\documentclass{ws-ijbc}

\usepackage{ws-rotating}     
\usepackage{lscape}

\newcommand{\sz}{\scriptsize}

\begin{document}

\catchline{}{}{}{}{} 

\markboth{Clara Granell {\it et al.}}{Unsupervised clustering analysis: a multiscale complex networks approach}

\title{UNSUPERVISED CLUSTERING ANALYSIS: A MULTISCALE COMPLEX NETWORKS APPROACH}

\author{CLARA GRANELL}
\address{Department d'Enginyeria Inform\`{a}tica i Matem\`{a}tiques, Universitat Rovira i Virgili\\
Av. Pa\"isos Catalans 26 Tarragona, 43007 Catalonia, Spain\\
clara.granell@urv.cat}

\author{SERGIO G\'OMEZ}
\address{Department d'Enginyeria Inform\`{a}tica i Matem\`{a}tiques, Universitat Rovira i Virgili\\
Av. Pa\"isos Catalans 26 Tarragona, 43007 Catalonia, Spain\\
sergio.gomez@urv.cat}

\author{ALEX ARENAS}
\address{Department d'Enginyeria Inform\`{a}tica i Matem\`{a}tiques, Universitat Rovira i Virgili\\
Av. Pa\"isos Catalans 26 Tarragona, 43007 Catalonia, Spain\\
alexandre.arenas@urv.cat}

\maketitle

\begin{history}
\received{(to be inserted by publisher)}
\end{history}

\begin{abstract}
Unsupervised clustering, also known as natural clustering, stands for the classification of data according to their similarities. Here we study this problem from the perspective of complex networks. Mapping the description of data similarities to graphs, we propose to extend two multiresolution modularity based algorithms to the finding of modules (clusters) in general data sets producing a multiscales' solution. We show the performance of these reported algorithms to the classification of a standard benchmark of data clustering and compare their performance.
\end{abstract}


\section{Introduction}

\noindent
The problem of unsupervised data clustering consists in classifying elements so that two data points
belonging to the same cluster are more similar between them than with elements in a different cluster.
An element, or pattern, is a vector of features (usually understood as a point in a multidimensional space)
that describes the item we wish to classify. The goal of the process of data clustering is to organize these
patterns finding a partition of the sample according to natural classes that are present in it. Data clustering has been the subject of interest in many disciplines where the mining of raw information is crucial to understand some phenomenon or gain insight into a system. It has applications in several fields such as pattern recognition, astronomic classification, biological taxonomy, marketing, and more \cite{clust_app}.

The methodology used to obtain the clusters from the raw data is as follows: First of all, a representation of the patterns has to be chosen, and also a feature selection or extraction is performed. Feature selection means choosing, from all the available features, those that will make easier the process of clustering, leaving the redundant, correlated and less informative features out of the analysis. On the other hand, feature extraction consists in transforming the original data set to a new one containing only the most relevant information. This first step is very important, as the result of the clustering often depends directly on its quality. Secondly, the similarity or dissimilarity between each pair of patterns has to be computed, which is often done by defining a measure of distance. The result of this step is the similarity matrix, which using the mapping to complex networks can be understood as a graph, where each node is a pattern and the links are the similarities between them \cite{rev_dc}. Finally, the main step of the process, the grouping (or clustering) algorithm, which will decompose the similarity matrix and return the groups of data.

The problem of clustering is inherently ill-posed, i.e. any data set can be clustered in drastically different ways, with no clear criterion for preferring one clustering over another. In particular, in the case of unsupervised approaches, a satisfactory clustering of data depends on the desired resolution which determines the number of clusters and their size. For example, $k$-means clustering fixes a priori the number of groups ($k$), which implies indeed a certain resolution of the clustering. Other algorithms such as hierarchical clustering \cite{clustering} group the patterns extending the measure of distance between them to distances between clusters of patterns. This process generates a complete dendrogram. Cutting the dendrogram at different heights we obtain different partitions of the data, all them hierarchically nested. In this situation the following question arises: To what resolution should one look at the data to find a scientific meaning in the classification? We claim that the answer to this question is totally dependent on the final purpose of the classification process, and that the concept of best solution should be reconsidered. Different partitions will be representative of properties of the data at different scales and then all of them are worth to be studied.

In this work we perform a comparison between two different multiresolution algorithms, used in the field of complex networks to detect community structure, applied to the problem of data clustering. We also compare our results with a hierarchical clustering (HC) algorithm. In contrast with hierarchical clustering the multiresolution methods are not necessarily hierarchical.
The first algorithm is the multiresolution static screening of the topology of the network, based on the introduction of a control parameter in the resolution of modularity \cite{njp08} (AFG method), proposed by the authors. The second one is a multiresolution dynamic screening of the network structure using a method, inspired in the Potts model, proposed by \cite{rb} (RB method). Both algorithms show to be competitive with classical clustering methods in the classification of the Iris data set.

\section{The complex networks approach}

\noindent
Complex networks are graphs representative of the intricate connections between elements in many natural and artificial systems \cite{strogatz,havlin,barabasi}, whose description in terms of statistical properties has been largely developed in the curse for a universal classification of them. However, when the networks are locally analyzed some characteristics that become partially hidden in the statistical description emerge. The most relevant perhaps is the discovery in many of them of {\em community structure}, meaning the existence of densely (or strongly) connected groups of nodes, with sparse (or weak) connections between them \cite{firstnewman}.

The study of the community structure  helps to elucidate the organization of the networks and, eventually, could be related to the functionality of groups of nodes \cite{amaral}. The most successful solutions to the community detection problem, in terms of accuracy, are those based in the optimization of a quality function called {\em modularity} proposed by Newman and Girvan \cite{newgirvan} that allows the comparison of different partitioning of the network. Given a network partitioned into communities, being $C_i$ the community to which node $i$ is assigned, the mathematical definition of modularity is expressed in terms of the weighted adjacency matrix $w_{ij}$, that represents the value of the weight in the link between nodes $i$ and $j$, this weight would be $0$ if no link existed, and the strengths $w_i=\sum_j w_{ij}$ as \cite{newanaly}
\begin{equation}
Q=\frac{1}{2w}\sum_i \sum_j \left(w_{ij}-\frac{w_i w_j}{2w}\right)\delta(C_i,C_j)\,,
\label{QW}
\end{equation}
\noindent where the Kronecker delta function $\delta(C_i,C_j)$ takes the value 1 if node $i$ and $j$ are into the same community,  0 otherwise, and the total strength is $2w=\sum_i w_i$.
The modularity of a given partition is then the probability of having  edges falling within groups in the network minus the expected probability in an equivalent (null case) network with the same number of nodes, and edges placed at random preserving the nodes' strength. The larger the modularity the best the partitioning is, cause more deviates from the null case. Note that the optimization of the modularity cannot be performed by exhaustive search since the number of different partitions is equal to the Bell or exponential numbers \cite{bell}, which grow at least exponentially in the number of nodes $N$. Indeed, optimization of modularity is a NP-hard (Non-deterministic Polynomial-time hard) problem \cite{brandes}. Several authors have attacked the problem, with considerable success, by proposing different optimization heuristics \cite{newfast, clauset, rogernat, duch, pujol, newspect}, see \cite{fortunato_rev} for a review.

Maximizing modularity one obtains the ``best" partition of the network into communities. This partition represents an intermediate topological scale of organization, or mesoscale, that in many cases has been shown to coincide with known information about subdivisions in the network \cite{newgirvan,jstat}. However, recently, it has been pointed out that the optimization of the modularity has a characteristic scale related to the number of links in the network, which delimits the resolution beyond which no separation into smaller groups can be obtained when optimizing modularity, even though these smaller partitions, and then different levels of description, are plausible to exist from direct observation \cite{fortunato}. The problem seems then that modularity, as it has been prescribed, does not have access to these other levels of description. The reason for this is that the topological scale at which we have access by maximizing modularity has a topological resolution limit.

We proposed a method that allows the full screening of the topological structure at any resolution level using the original formulation and semantics of modularity, overcoming then the resolution limit \cite{njp08}. Our aim is to take advantage of this method to analyze real data sets in terms of clustering. In contrast with the solution proposed in \cite{angelini} to find the correct clustering using modularity, here we present a multiple scale method based on the optimization of modularity as well.
The mathematical form of our prescription is given by $Q_{\mbox{\sz AFG}}(r)=Q [w_{ij}\leftarrow w_{ij}+ r \delta_{ij}]$ where $r$ (resistance) is the parameter controlling the resolution of the partitions we want to find, and $w_{ij}+ r \delta_{ij}$ is the new weights' matrix after adding a self-loop with value $r$ to each node. The definition of $Q_{\mbox{\sz AFG}}$ does preserve the original semantics of modularity.

A different approach was proposed by Reichardt and Bornholdt \cite{rb}, in their work every node can be understood as a dynamical system of oscillators of q-states (usually known as Potts' model), and the partition in modules is equivalent to the ground state of the mentioned dynamical system. Indeed, the authors made a very interesting connection with the statistical mechanics of the Potts model and modularity. Moreover, although the finding of the resolution limit was discovered later, the RB method already solved this problem by the tuning of a parameter, as pointed out in \cite{kumpula}.
The result is that the ground state of the system corresponding to the minimum of its Hamiltonian can be written as

\begin{equation}
Q_{\mbox{\sz RB}}(\gamma)=\frac{1}{2w}\sum_i \sum_j \left( w_{ij}-\gamma \frac{w_i w_j}{2w}\right)\delta(C_i,C_j)\,,
\label{QWRB}
\end{equation}

\noindent where $\gamma$ is the resolution control parameter in this case. Note that the original $Q$ corresponds to $\gamma=1$ where other values are different quality functions characterized by a weight the null model term.

To screen the whole spectrum of resolution levels of the topological structure of any given network, we must determine the values of $r_{\min}$ and $r_{\max}$ for the AFG model, and the $\gamma_{\min}$ and $\gamma_{\max}$ for the RB model, which will make the network to appear as an unique module or as a set of as many modules as nodes in the network.The mathematical determination of these limits is discussed in the Appendix for the most general case of directed and signed networks. The screening of the mesoscale is done by optimizing modularity $Q_{\mbox{\sz AFG}}(r)$, and optimizing modularity in the $Q_{\mbox{\sz RB}}(\gamma)$, for the different values of $r$ and $\gamma$ respectively.

\section{Results}

\begin{figure}
  \begin{center}
    \mbox{\includegraphics*[width=.650\textwidth]{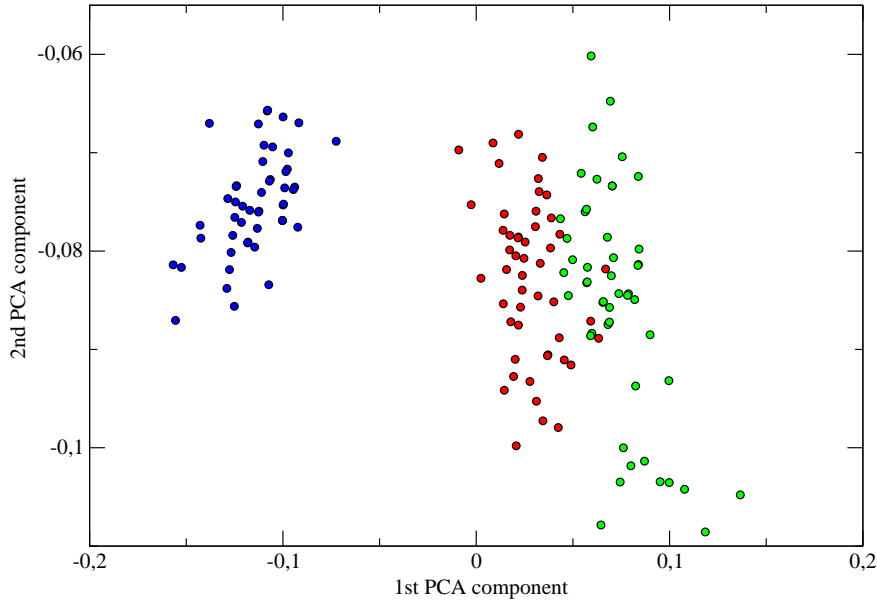}}
  \end{center}
  \caption{Two principal components of the PCA analysis on the Iris data set. Colors correspondence is: {\em setosa}-blue, {\em versicolor}-red, and {\em virginica}-green. While {\em setosa} is clearly linearly separable, the other two species are not.}
  \label{pca}
\end{figure}

\noindent
To show the ability of multiresolution community detection methods to solve the problem of unsupervised data clustering, we have chosen to study the classical benchmark of the Iris data set. This dataset, presented by Sir R.A. Fisher in 1936, consists of 50 samples from each of three species of Iris flowers ({\em Iris setosa}, {\em Iris versicolor} and {\em Iris virginica}). We know the petal length, petal width, sepal length and sepal width from each sample. For the moment, we will ignore the species information and we will cluster the data using only the raw measurements as in \cite{fisher}. When this is done, a comparison between the real classification and the obtained clusters can be made, in order to evaluate its quality.

\begin{figure}[!tpb]
  \begin{center}
  \begin{tabular}{ll}
    \begin{tabular}{l}
    a) Multidendrogram
    \\ \\
    \mbox{\includegraphics*[width=.46\textwidth]{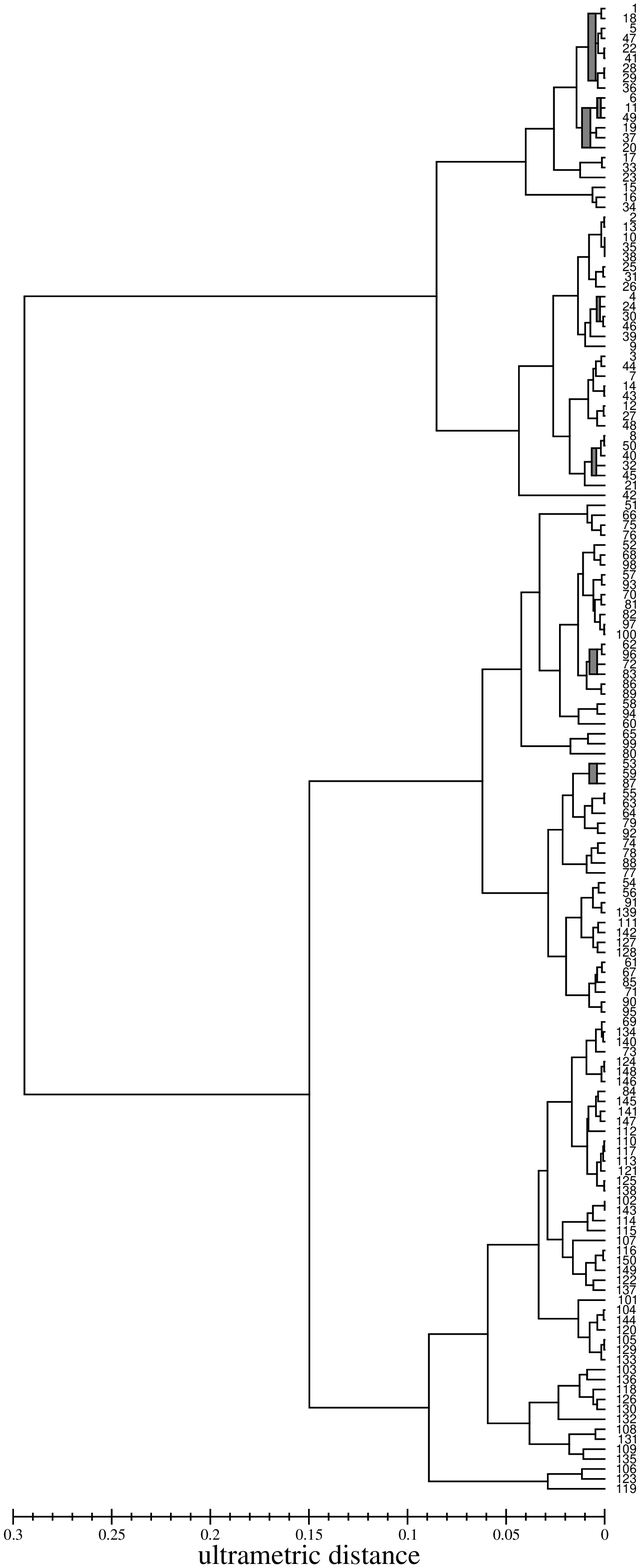}}
    \end{tabular}
  &
    \begin{tabular}{l}
    b) AFC mesoscales
    \\ \\
    \mbox{\includegraphics*[width=.45\textwidth]{fig2b.eps}}
    \\ \\
    c) RB mesoscales
    \\ \\
    \mbox{\includegraphics*[width=.45\textwidth]{fig2c.eps}}
    \\ \\
    d) HC mesoscales
    \\ \\
    \mbox{\includegraphics*[width=.45\textwidth]{fig2d.eps}}
    \\
    \mbox{\rule{0pt}{1pt}}
    \end{tabular}
  \end{tabular}
  \end{center}
  \caption{Mesoscales of the Iris data set, showing the number of clusters as a function of the resolution parameter: a) Complete linkage multidendrogram; b) AFG mesoscales; c) RB mesoscales; d) HC mesoscales from the previous multidendrogram.}
\label{dendo}
\end{figure}
Following the steps of data clustering explained above, we first perform a principal component analysis of the four features that form each pattern, and choose to work with the two principal components corresponding to the largest part of the data variance. In Fig.~\ref{pca} a representation of these two components is shown. Based on these two variables, we build up a similarity matrix from the euclidean distances between patterns components with respect to the average distance in this space. For any pair of flowers $i$ and $j$, we define the similarity $s_{ij}={\bar d} - \| x^{i}-x^{j} \|)$, where ${\bar d}$ stands for the average distance of the set, and $\| \cdot \|$ is the euclidean distance between the feature vectors of each flower. The resulting similarity matrix is interpreted as a weighted network whose communities will, in principle, reproduce the right clustering of the data. Note that this matrix has positive and negative links, and that modularity should account for this signed values, see Appendix.

We present the comparison of the results obtained using the algorithms described above, and also compare with the solution obtained applying a classical hierarchical clustering technique, see Fig.~\ref{dendo}.

In particular, we constructed the hierarchical clustering using complete linkage, where the distance between groups is defined as the distance between the most distant pair of individuals, one from each group. In other words, the distance between two clusters is given by the value of the longest link between the clusters. At each stage of hierarchical clustering, the clusters at minimum distance are merged. Moreover, instead of using the standard pair-group hierarchical clustering approach, we take advantage of a recent development by some of the authors \cite{multidendro} that allows to solve the non-uniqueness problem when there are tied distances during the agglomeration process (code available at \cite{wwwmultidendro}). The result, known as a {\em multidendrogram}, is presented in Fig.~\ref{dendo}a. We plot the tag number of each specimen at the leaves of the tree. The analysis of the multidendrogram can be performed as follows: starting from the root of the tree, we can compute the distances between different partitions of the data and analyze each of them separately.

The comparison between the three methods can be done by computing the multiple scales of the topology in terms of community structure, screening the values of $r$ in the AFG method, the values of $\gamma$ in the RB method, and the distances in the dendrogram. In Fig.~\ref{dendo}b we present the whole mesoscale for the AFG method, we observe the persistence of the partition in two clusters, and the partition in three clusters as the more representatives of the mesoscale. In Fig.~\ref{dendo}c we present a portion of the mesoscale for the RB method, again the last observation holds for this method, however, the variations of $\gamma$ do not ensure a monotonic behavior of the number of clusters as a function of $\gamma$ (see Appendix for details). Finally, we plot the mesoscale in terms of distances in the dendrogram, see Fig.~\ref{dendo}d. The hierarchical clustering approach defines also two main resolution levels corresponding to two and three clusters partitions, respectively. The fact that the partition that divides the data in two communities is always the most relevant in any of the used methods corresponds to the true partition of the Iris data set in two linearly separable sets.

\begin{figure}
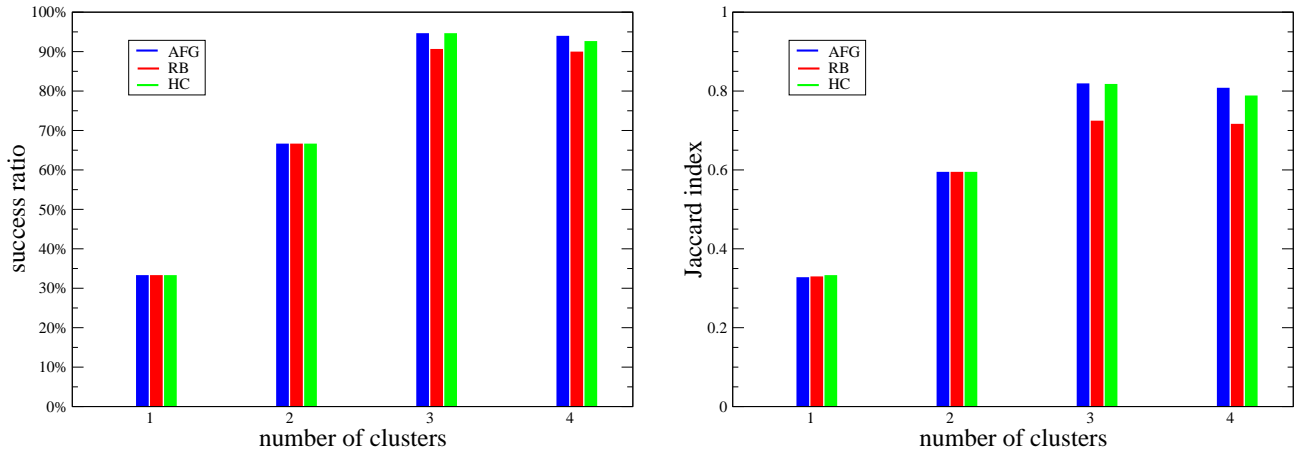

  \begin{center}
  \begin{tabular}{ll}
    \mbox{\includegraphics*[width=.47\textwidth]{fig3a.eps}}&
    \mbox{\includegraphics*[width=.47\textwidth]{fig3b.eps}}
  \end{tabular}
  \end{center}
  \caption{Comparison between the three methods used in the classification of the Iris data set. Two measures are used: the success ratio (left) and the Jaccard index (right). Only the partitions with highest performance and less than five clusters are shown.}
\label{success}
\end{figure}

We define two measures to make the comparison between the different methods, centering our attention in the most relevant partitions in terms of the scale length, see Fig.~\ref{success}. The first measure is the success, which is computed as the percentage of correctly classified nodes when comparing the partition obtained with the original classification taxonomy made by biologists using more features of the flowers. In this case and for the partition in three clusters, both HC and AFG methods achieve a $94,67\%$ of success, corresponding to a mismatch of eight flowers in total. The RB method obtains a success of $90,67\%$ in this case. The second measure we contemplate is the Jaccard index presented in \cite{jaccard}, which is the fraction of pairs of patterns in the same cluster in one partition which are also in the same cluster in the other partition. The larger the fraction of same cluster co-ocurrences, the better the quality of the agreement. In Fig.~\ref{success}(right) we observe that the best classification in three clusters is performed by the AFG method by a slight difference ($0.8194$ the AFC method versus $0.8180$ the HC).

\section{Conclusions}

\noindent
We have presented the adaptation and performance of two multiresolution methods, for the determination of the community structure in networks, to the problem of unsupervised data clustering. We focus on the determination of groups in the similarity matrix using modularity as the quality function. We have analytically computed the two limiting cases for the AFG method corresponding to the classification of the set as a unique cluster to the classification of every data point as a single cluster. The results on the classical Iris data set are competitive with classical unsupervised clustering techniques. These results are encouraging, and point out that the mapping of clustering problems to networks' structural analysis is a field worth to be explored.

\nonumsection{Acknowledgments}
We acknowledge support from the Spanish Ministry of Science and Technology FIS2009-13730-C02-02 and the Generalitat de Catalunya SGR-00838-2009.

\appendix{Determination of AFG mesoscales boundaries}

\noindent
The generalization of modularity Eq.~(\ref{QW}) for undirected weighted signed networks (see \cite{signed}) is
\begin{equation}
  Q = \frac{1}{2w^{+} + 2w^{-}} \sum_i \sum_j
        \left[ w_{ij} - \left(
          \frac{w_i^{+} w_j^{+}}{2w^{+}} - \frac{w_i^{-} w_j^{-}}{2w^{-}}
        \right) \right]
        \times \delta(C_i,C_j)\,.
  \label{QWS}
\end{equation}
where
\begin{eqnarray}
  w_i^{+} & = & \sum_{j,w_{ij}>0} w_{ij}\,,
  \\
  w_i^{-} & = & \sum_{j,w_{ij}<0} |w_{ij}|\,,
\end{eqnarray}
are the positive and negative strengths of node $i$, and
\begin{eqnarray}
  2w^{+} & = & \sum_i w_i^{+}\,,
  \\
  2w^{-} & = & \sum_i w_i^{-}\,,
\end{eqnarray}
the positive and negative total strengths respectively. Please note that
these four strengths are defined to be non-negative.

To simplify the notation, we make use of the modularity matrix
\begin{equation}
  B_{ij} = w_{ij} - \left( \frac{w_i^{+} w_j^{+}}{2w^{+}} - \frac{w_i^{-} w_j^{-}}{2w^{-}} \right)\,,
  \label{Bij}
\end{equation}
therefore
\begin{equation}
  Q = \frac{1}{2w^{+} + 2w^{-}} \sum_{i=1}^N \sum_{j=1}^N B_{ij} \delta(C_i,C_j)\,.
\end{equation}

Following \cite{njp08}, the analysis of the mesoscale is performed with the addition of a common self-loop to all the nodes in the network. The boundaries of the mesoscale are the {\em macroscale}, a partition in which all nodes belong to the same community, and the {\em microscale}, a partition in which each node is isolated in its own community. The determination of these boundaries is equivalent to finding two values of the self-loops, $r_{\min}$ and $r_{\max}$, for which the maximum of modularity $Q_{\mbox{\sz AFG}}(r)$ is achieved at the macroscale and microscale respectively. The solution is quite simple: if all the non-diagonal terms of the modularity matrix are positive or zero, modularity is optimized at the macroscale, and if they are negative, it is optimized at the microscale. Diagonal terms are irrelevant since $\delta(C_i,C_i)=1$ for all nodes.

If we introduce a positive self-loop $r^{+}$, the modularity matrix becomes
\begin{equation}
  B^{\mbox{\sz AFG}}_{ij}(r^{+}) = w_{ij} + r^{+}\delta_{ij} - \left(
    \frac{(w_i^{+}+r^{+}) (w_j^{+}+r^{+})}{2w^{+}+Nr^{+}} - \frac{w_i^{-} w_j^{-}}{2w^{-}}
    \right)\,,
  \label{Brp}
\end{equation}
and with a negative self-loop $-r^{-}$
\begin{equation}
  B^{\mbox{\sz AFG}}_{ij}(-r^{-}) = w_{ij} - r^{-}\delta_{ij} - \left(
    \frac{w_i^{+} w_j^{+}}{2w^{+}} - \frac{(w_i^{-} + r^{-}) (w_j^{-} + r^{-})}{2w^{-}+Nr^{-}}
    \right)\,.
  \label{Brm}
\end{equation}

The existence of $r_{\max}$ is straightforward, since $B^{\mbox{\sz AFG}}_{ij}(r^{+}) \sim - r^{+} < 0$ for large enough $r^{+}$ and $i\neq j$. Its determination is just an exercise of solving the system of inequations $B^{\mbox{\sz AFG}}_{ij}(r^{+}) \leq 0$ for $i<j$, and taking the smallest solution as $r_{\max}$. More precisely,
\begin{equation}
  r_{\max} = \max_{\substack{i < j \\ D_{ij}^2 \ge 4E_{ij}}}
    \left(-\frac{D_{ij}}{2} + \frac{1}{2}\sqrt{D_{ij}^2 - 4E_{ij}}\,\right)\,,
\end{equation}
where
\begin{eqnarray}
  D_{ij} & = & w_i^{+} + w_j^{+} - N\left(w_{ij} + \frac{w_i^{-} w_j^{-}}{2w^{-}}\right)\,,
  \\
  E_{ij} & = & w_i^{+} w_j^{+} - 2w^{+}\left(w_{ij} + \frac{w_i^{-} w_j^{-}}{2w^{-}}\right)\,.
\end{eqnarray}

In the same way, $B^{\mbox{\sz AFG}}_{ij}(-r^{-}) \sim r^{-} > 0$ proves the existence of $r_{\min}$, and it is calculated by solving $B^{\mbox{\sz AFG}}_{ij}(-r^{-}) \geq 0$ for $i<j$, and taking the largest solution as $r_{\min}$, i.e.
\begin{equation}
  r_{\min} = -\max_{\substack{i < j \\ D_{ij}^2 \ge 4E_{ij}}}
    \left(-\frac{D_{ij}}{2} + \frac{1}{2}\sqrt{D_{ij}^2 - 4E_{ij}}\,\right)\,,
\end{equation}
where
\begin{eqnarray}
  D_{ij} & = & w_i^{-} + w_j^{-} + N\left(w_{ij} - \frac{w_i^{+} w_j^{+}}{2w^{+}}\right)\,,
  \\
  E_{ij} & = & w_i^{-} w_j^{-} + 2w^{-}\left(w_{ij} - \frac{w_i^{+} w_j^{+}}{2w^{+}}\right)\,.
\end{eqnarray}

When the network is directed, the analysis of the AFG mesoscale is exactly the same, but with the substitutions
\begin{eqnarray}
  w_{i}^{\pm} & \rightarrow &  w_{i}^{\pm,\mbox{\scriptsize out}} = \sum_{k} w_{ik}\,,
  \\
  w_{j}^{\pm} & \rightarrow &  w_{j}^{\pm,\mbox{\scriptsize in}} = \sum_{k} w_{kj}\,,
  \\
  D_{ij} & \rightarrow & \frac{1}{2} (D_{ij} + D_{ji})\,,
  \\
  E_{ij} & \rightarrow & \frac{1}{2} (E_{ij} + E_{ji})\,.
\end{eqnarray}

\appendix{Boundaries of RB mesoscales}

\noindent
In the RB formulation of mesoscales, a parameter $\gamma$ is introduced in front of the null-case term to weight its relative importance against the real network, i.e.
\begin{equation}
  B^{\mbox{\sz RB}}_{ij}(\gamma) = w_{ij} - \gamma \left(
    \frac{w_i^{+} w_j^{+}}{2w^{+}} - \frac{w_i^{-} w_j^{-}}{2w^{-}}
    \right)\,.
  \label{BRB}
\end{equation}
It is also possible to have different parameters for the positive and negative null-case terms as in \cite{traag}, however this leads to a bidimensional analysis of the mesoscales, which is almost unaffordable for most real networks. Thus, we will focus on the single-parameter RB modularity matrix Eq.~(\ref{BRB}).

Without negative weights, the macroscale is recovered at $\gamma_{\min}=0$, and the microscale at the $\gamma_{\max}$ which makes all modularity terms negative. The existence of $\gamma_{\max}$ is guaranteed by the fact that all null-case terms are positive. However, the addition of negative weights makes it possible to have both positive and negative null-case terms, which does not allow to ensure the recovery of macro and microscale. Therefore, RB signed modularity may not cover the whole mesoscale.
\noindent
\begin{figure}[!tpb]
  \begin{center}
    \mbox{\includegraphics*[width=.70\textwidth]{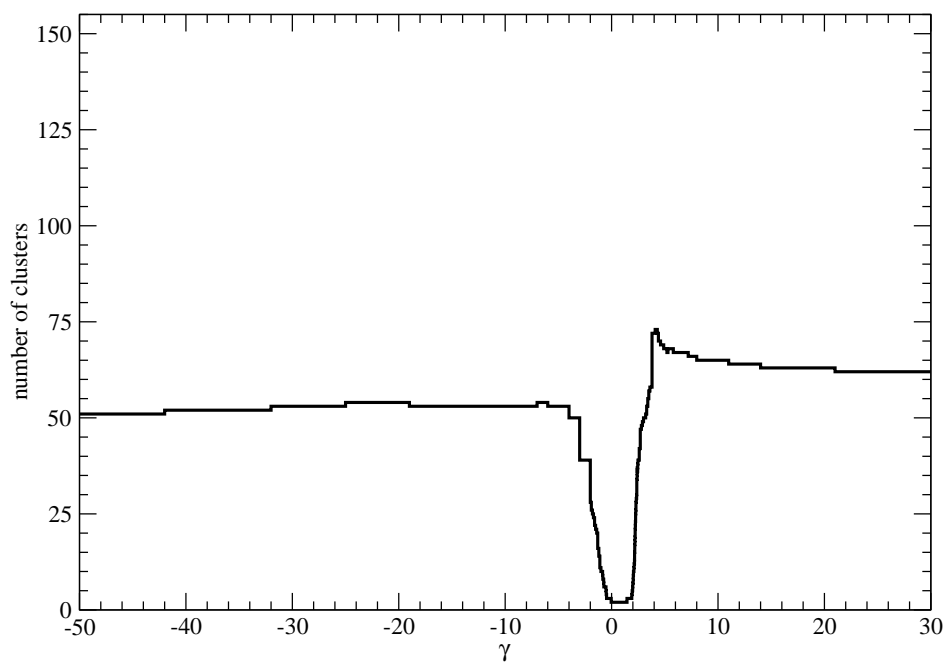}}
  \end{center}
  \caption{Expanded Iris data set RB mesoscales analysis.}
\label{RBcomplete}
\end{figure}
This is experimentally confirmed in Fig.~\ref{RBcomplete} for the Iris data set, where a larger interval of the $\gamma$ parameter has been analyzed. While Fig.~\ref{dendo}c only shows the useful part of the mesoscales range, where the number of clusters goes from 2 to 73 ($\gamma \in [0.0,4.2]$), in Fig.~\ref{RBcomplete} it is shown the inability of RB to find the macroscale (microscale) for lower (larger) values of $\gamma$.

\bibliographystyle{ws-ijbc}
\bibliography{cga}

\begin{thebibliography}{31}
\newcommand{\enquote}[1]{``#1''}
\providecommand{\natexlab}[1]{#1}
\providecommand{\url}[1]{\texttt{#1}}
\providecommand{\urlprefix}{URL }
\expandafter\ifx\csname urlstyle\endcsname\relax
  \providecommand{\doi}[1]{doi:\discretionary{}{}{}#1}\else
  \providecommand{\doi}{doi:\discretionary{}{}{}\begingroup
  \urlstyle{rm}\Url}\fi

\bibitem[{Angelini \emph{et~al.}(2007)Angelini, Marinazzo, Pellicoro \&
  Stramaglia}]{angelini}
Angelini, L., Marinazzo, D., Pellicoro, M. \& Stramaglia, S. [2007]
  \enquote{Natural clustering: the modularity approach,} \emph{J. Stat. Mech.}
  .

\bibitem[{Arenas \emph{et~al.}(2008)Arenas, Fern\'andez \& G\'omez}]{njp08}
Arenas, A., Fern\'andez, A. \& G\'omez, S. [2008] \enquote{Multiple resolution
  of the modular structure of complex networks,} \emph{New Journal of Physics}
  \textbf{10}.

\bibitem[{Barab{\'a}si(2005)}]{barabasi}
Barab{\'a}si, A.-L. [2005] \enquote{Network theory-the emergence of the
  creative enterprise,} \emph{Science} \textbf{433}.

\bibitem[{Bell(1934)}]{bell}
Bell, E.~T. [1934] \enquote{Exponential numbers,} \emph{Amer. Math. Monthly}
  \textbf{41},  411.

\bibitem[{Brandes \emph{et~al.}(2008)Brandes, Delling, Gaertler, Goerke,
  Hoefer, Nikoloski \& Wagner}]{brandes}
Brandes, U., Delling, D., Gaertler, M., Goerke, R., Hoefer, M., Nikoloski, Z.
  \& Wagner, D. [2008] \enquote{On modularity clustering,} \emph{IEEE Trans.
  Knowl. Data Eng.} \textbf{20},  172.

\bibitem[{Clauset \emph{et~al.}(1994)Clauset, Newman \& Moore}]{clauset}
Clauset, A., Newman, M. E.~J. \& Moore, C. [1994] \enquote{Finding community
  structure in very large networks,} \emph{Phys. Rev. E} \textbf{70},  066111.

\bibitem[{Danon \emph{et~al.}(2005)Danon, D\'iaz-Guilera, Duch \&
  Arenas}]{jstat}
Danon, L., D\'iaz-Guilera, A., Duch, J. \& Arenas, A. [2005] \enquote{Comparing
  community structure identification,} \emph{J. Stat. Mech.} ,  P09008.

\bibitem[{Duch \& Arenas(2005)}]{duch}
Duch, J. \& Arenas, A. [2005] \enquote{Community identification using extremal
  optimization,} \emph{Phys. Rev. E} \textbf{72}.

\bibitem[{Fern\'andez \& G\'omez(2008)}]{multidendro}
Fern\'andez, A. \& G\'omez, S. [2008] \enquote{Solving non-uniqueness in
  agglomerative hierarchical clustering using multidendrograms,} \emph{Journal
  of Classification} \textbf{25},  43.

\bibitem[{Fisher(1936)}]{fisher}
Fisher, R.~A. [1936] \enquote{The use of multiple measurements in taxonomic
  problems,} \emph{Annals of Eugenics} \textbf{7},  179.

\bibitem[{Fortunato(2010)}]{fortunato_rev}
Fortunato, S. [2010] \enquote{Community detection in graphs,} \emph{Phys. Rep.}
  \textbf{486},  75.

\bibitem[{Fortunato \& Barth{\'e}lemy(2007)}]{fortunato}
Fortunato, S. \& Barth{\'e}lemy, M. [2007] \enquote{Resolution limit in
  community detection,} \emph{Proc. Natl. Acad. Sci. USA} \textbf{104},  36.

\bibitem[{Gan \emph{et~al.}(2007)Gan, Ma \& Wu}]{clust_app}
Gan, G., Ma, C. \& Wu, J. [2007] \emph{Data Clustering: Theory, Algorithms, and
  Applications}, Series on Statistics and Applied Probability (ASA-SIAM).

\bibitem[{Girvan \& Newman(2002)}]{firstnewman}
Girvan, M. \& Newman, M. E.~J. [2002] \enquote{Community structure in social
  and biological networks,} \emph{Proc. Natl. Acad. Sci. USA} \textbf{99},
  7821.

\bibitem[{G\'omez(2010)}]{wwwmultidendro}
G\'omez, S. [2010]
  \urlprefix\url{http://deim.urv.cat/~sgomez/multidendrograms.php}.

\bibitem[{G\'omez \emph{et~al.}(2009)G\'omez, Jensen \& Arenas}]{signed}
G\'omez, S., Jensen, P. \& Arenas, A. [2009] \enquote{Analysis of community
  structure in networks of correlated data,} \emph{Phys. Rev. E} \textbf{80},
  016114.

\bibitem[{Guimer{\`a} \& Amaral(2005{\natexlab{a}})}]{rogernat}
Guimer{\`a}, R. \& Amaral, L. A.~N. [2005{\natexlab{a}}] \enquote{Cartography
  of complex networks: modules and universal roles,} \emph{J. Stat. Mech.} .

\bibitem[{Guimer{\`a} \& Amaral(2005{\natexlab{b}})}]{amaral}
Guimer{\`a}, R. \& Amaral, L. A.~N. [2005{\natexlab{b}}] \enquote{Functional
  cartography of complex metabolic networks,} \emph{Nature} \textbf{433},  895.

\bibitem[{Jaccard(1912)}]{jaccard}
Jaccard, P. [1912] \enquote{The distribution of flora in the alpine zone,}
  \emph{The New Phytologist} \textbf{11}.

\bibitem[{Jain \emph{et~al.}(1999)Jain, Murty \& Flynn}]{rev_dc}
Jain, A.~K., Murty, M.~N. \& Flynn, P.~J. [1999] \enquote{Data clustering: A
  review,} \emph{ACM Comp. Surv.} \textbf{31}.

\bibitem[{Kaufman \& Rousseeuw(2005)}]{clustering}
Kaufman, L. \& Rousseeuw, P.~J. [2005] \emph{Finding Groups in Data: An
  Introduction to Cluster Analysis} (Wiley).

\bibitem[{Kumpula \emph{et~al.}(2007)Kumpula, Saramaki, Kaski \&
  Kertesz}]{kumpula}
Kumpula, J.~M., Saramaki, J., Kaski, K. \& Kertesz, J. [2007] \enquote{Limited
  resolution and multiresolution methods in complex network community
  detection,} \emph{Fluctuation Noise Letters} ,  L209.

\bibitem[{Newman(2004{\natexlab{a}})}]{newanaly}
Newman, M. E.~J. [2004{\natexlab{a}}] \enquote{Analysis of weighted networks,}
  \emph{Phys. Rev. E} \textbf{70},  056131.

\bibitem[{Newman(2004{\natexlab{b}})}]{newfast}
Newman, M. E.~J. [2004{\natexlab{b}}] \enquote{Fast algorithm for detecting
  community structure in networks,} \emph{Phys. Rev. E} \textbf{69},  066133.

\bibitem[{Newman(2006)}]{newspect}
Newman, M. E.~J. [2006] \enquote{Modularity and community structure in
  networks,} \emph{Proc. Natl. Acad. Sci. USA} \textbf{103},  8577.

\bibitem[{Newman \& Girvan(2004)}]{newgirvan}
Newman, M. E.~J. \& Girvan, M. [2004] \enquote{Finding and evaluating community
  structure in networks,} \emph{Phys. Rev. E} \textbf{69},  026113.

\bibitem[{Pujol \emph{et~al.}(2006)Pujol, B{\'e}jar \& Delgado}]{pujol}
Pujol, J.~M., B{\'e}jar, J. \& Delgado, J. [2006] \enquote{Clustering algorithm
  for determining community structure in large networks,} \emph{Phys. Rev. E}
  \textbf{74}.

\bibitem[{Reichardt \& Bornholdt(2004)}]{rb}
Reichardt, J. \& Bornholdt, S. [2004] \enquote{Detecting fuzzy community
  structures in complex networks with a potts model,} \emph{Phys. Rev. Lett.}
  \textbf{93}.

\bibitem[{Song \emph{et~al.}(2005)Song, Havlin \& Makse}]{havlin}
Song, C.~M., Havlin, S. \& Makse, H.~A. [2005] \enquote{Self-similarity of
  complex networks,} \emph{Nature} \textbf{433}.

\bibitem[{Strogatz(2001)}]{strogatz}
Strogatz, S.~H. [2001] \enquote{Exploring complex networks,} \emph{Nature}
  \textbf{410}.

\bibitem[{Traag \& Bruggeman(2009)}]{traag}
Traag, V. \& Bruggeman, J. [2009] \enquote{Community detection in networks with
  positive and negative links,} \emph{Phys. Rev. E} \textbf{80}.

\end{thebibliography}

\end{document}